# Pressure-Induced Irreversible Disorder in β'-$Mn_3(PO_4)_2$: A High-Pressure X-ray Diffraction and Density-Functional Theory Study

Ana Melissa P. Brito[1], Neha Bura[2], Pablo Botella[2], Robert Oliva[3], Alanna Khésley L. da Costa[1], Fabiana V. da Motta[1], Mauricio R. D. Bomio[1], Alfonso Muñoz[4], João Elias Rodrigues[5], Daniel Errandonea[2,*]

[1]LSQM—Laboratory of Chemical Synthesis of Materials—Department of Materials Engineering, Federal University of Rio Grande do Norte, Natal, Rio Grande do Norte, Brazil

[2]Departamento de Fisica Aplicada-ICMUV, MALTA Consolider Team, Universitat de Valencia, Av. Dr Moliner 50, 46100 Burjassot, Valencia, Spain

[3]Geosciences Barcelona (geo3bcn-CSIC), C/ Lluís Solé i Sabarís s/n, 08028 Barcelona, Catalonia, Spain

[4]Departamento de Física, MALTA Consolider Team, Universidad de La Laguna, E-38200 San Cristóbal de La Laguna, Tenerife, Spain

[5]European Synchrotron Radiation Facility, Grenoble 38043, France

*Corresponding author: daniel.errandonea@uv.es

**Abstract:** The high-pressure structural behavior of β'-$Mn_3(PO_4)_2$ was investigated using synchrotron X-ray diffraction up to 20 GPa combined with density-functional theory calculations. At ambient conditions, β'-$Mn_3(PO_4)_2$ crystallizes in a monoclinic structure that exhibits strongly anisotropic compression. The pressure dependence of the unit-cell volume was described using a third-order Birch–Murnaghan equation of state, and the principal axes of compressibility were determined. Above 14.1 GPa, significant broadening and weakening of the diffraction peaks are attributed to the onset of irreversible pressure-induced structural disorder associated with the loss of long-range crystallographic order. The disordered state persists after decompression to ambient pressure, demonstrating the irreversible nature of the transformation. The calculations accurately reproduce the experimental compressional behavior and provide insights into the microscopic structural evolution under pressure. Compression is mainly accommodated through distortions of the Mn–O polyhedra, whereas the $PO_4$ tetrahedra behave as comparatively rigid units. Several initially penta-coordinated Mn sites progressively evolve toward octahedral coordination under compression, while selected $MnO_6$ polyhedra exhibit anomalous distortions and elastic softening preceding the onset of disorder. Elastic constant calculations further reveal that the crystalline phase becomes mechanically unstable near the experimentally observed transition pressure. The combined experimental and computational results suggest that the high-pressure response of β'-$Mn_3(PO_4)_2$ is influenced by the interplay between framework complexity, anisotropic polyhedral compressibility, and elastic instability, providing new insight into pressure-induced structural degradation in structurally complex phosphate frameworks.

## I. Introduction

High-pressure (HP) research has become a powerful and versatile approach for probing and tuning the structural, electronic, and magnetic properties of complex oxides.[1,2] Compression modifies interatomic distances and bonding interactions without introducing chemical disorder, enabling the exploration of structural stability, phase transitions, coordination changes, and emergent physical properties inaccessible at ambient conditions.[3] In transition-metal oxides, pressure can strongly influence the balance between competing crystal structures and electronic states, often leading to novel phases with distinct functional properties.[4,5] Consequently, HP studies provide valuable insight into the fundamental mechanisms governing structure–property relationships and are also relevant for applications in energy materials, catalysis, electronics, and geoscience.[6,7,8] In particular, compounds containing tetrahedral oxyanions such as vanadates and phosphates constitute an important class of materials because their open-framework structures exhibit a rich variety of compression mechanisms and pressure-induced transformations.[9,10]

Orthovanadates and related vanadate compounds have attracted considerable attention in high-pressure research due to their rich structural behavior and pressure-induced phase transitions.[10,11] Compression studies have revealed a wide variety of structural responses, including symmetry changes, amorphization, and coexistence of metastable phases, highlighting the remarkable flexibility of transition-metal vanadate frameworks.[10] In particular, pressure-driven structural transformations have been extensively investigated in several vanadates and orthovanadates, demonstrating the strong coupling between crystal structure, coordination environment, and external pressure.[10] For example, phase coexistence between different structural modifications has been reported in $Mn_3(VO_4)_2$ over a broad pressure range between 10 and 20 GPa and amorphization above 20 GPa,[12] while studies on $Ni_3(VO_4)_2$ and related compounds have revealed complex pressure-induced structural evolution.[13] Likewise, $Cu_3(VO_4)_2$ exhibits decomposition under compression, further confirming the sensitivity of vanadate frameworks to external pressure.

In contrast, orthophosphates have received comparatively less attention under high-pressure conditions despite their technological and geophysical relevance. The limited number of available studies nevertheless suggests that phosphates may exhibit equally complex structural behavior. For instance, $Co_3(PO_4)_2$ undergoes a pressure-induced phase transition near 3 GPa,[15] whereas $Zn_3(PO_4)_2$ loses long-range structural order at approximately 14 GPa.[16] In the case of $Mg_3(PO_4)_2$, a triclinic high-pressure phase has been synthesized under combined high-pressure and high-temperature conditions at 2.2 GPa and 1000 °C.[17] These observations indicate that phosphate frameworks can display a broad spectrum of structural responses under compression, ranging from crystalline phase transitions to pressure-induced disorder.

Among transition-metal phosphates, manganese phosphate, $Mn_3(PO_4)_2$, is particularly interesting because of its rich polymorphism.[18,19] Four different polymorphs have been reported, with the monoclinic β' phase being the thermodynamically most stable form at ambient conditions.[20] Previous experiments conducted at high-pressure and high-temperature conditions have demonstrated that the β' phase transforms into the γ phase at approximately 2.5 GPa and 600 °C,[18] highlighting the delicate energetic balance between competing crystal structures in this compound. The γ phase, while also monoclinic, features a significantly simpler and less dense crystal structure compared to β'-$Mn_3(PO_4)_2$. A comprehensive description can be found in Refs. 18 and 19. However, the behavior of manganese phosphate under purely hydrostatic compression remains poorly understood.

In this work, we investigate the HP structural behavior of β'-$Mn_3(PO_4)_2$ by combining synchrotron high-pressure X-ray diffraction (XRD) measurements with density-functional theory (DFT) calculations. Our results reveal that compression induces significant structural disorder in the material, providing new insight into the response of phosphate frameworks to external pressure. More broadly, the present work aims to clarify how framework topology, polyhedral compressibility, and elastic stability influence the compression mechanisms of structurally complex transition-metal phosphates and why closely related compounds can exhibit markedly different high-pressure responses.

## II. Methods and calculations

### a. Sample preparation

Monobasic ammonium phosphate [$(NH_4)H_2PO_4$] (98.5% purity; Sigma-Aldrich), manganese(II) nitrate tetrahydrate [$Mn(NO_3)_2 \cdot 4H_2O$] (98% purity; Alfa Aesar), and distilled water were used as precursor materials. All reagents were used as received without further purification. β'-$Mn_3(PO_4)_2$ was synthesized by a co-precipitation method, dissolving 1.17 g of $(NH_4)H_2PO_4$ and 2.56 g of $Mn(NO_3)_2 \cdot 4H_2O$ in 100 mL of distilled water under continuous stirring at room temperature for 1 h. Precipitation was induced by adding by adding ammonium hydroxide ($NH_4OH$, 30% purity; Synth) until the pH reached 5. The pinkish-white precipitate obtained was washed three times with distilled water and twice with ethanol. The sample was dried in an oven at 100 °C for 24 h and subsequently calcined in a muffle furnace at 700 °C for 5 h with a heating rate of 5 °C min$^{-1}$, yielding β'-$Mn_3(PO_4)_2$ powder.

### b. Sample characterization

Figure 1 shows a powder diffraction (XRD) pattern measured at ambient conditions for the synthesized sample, using a Malvern PANalytical X'Pert Pro diffractometer with Cu K$\alpha_1$ radiation ($\lambda$ = 1.5406 Å). The XRD results confirmed that the compound investigated corresponds to the β' phase of $Mn_3(PO_4)_2$. The crystal belongs to the monoclinic crystal system with space group $P2_1/c$ and contains 12 formula units per unit cell.[20] The unit-cell parameters obtained from the Rietveld refinement of the experimental XRD pattern are $a = 8.942(4)$ Å, $b = 10.043(5)$ Å, $c = 24.121(12)$ Å, and $\beta = 120.79(5)°$. These values are in good agreement with previously reported results.[20]

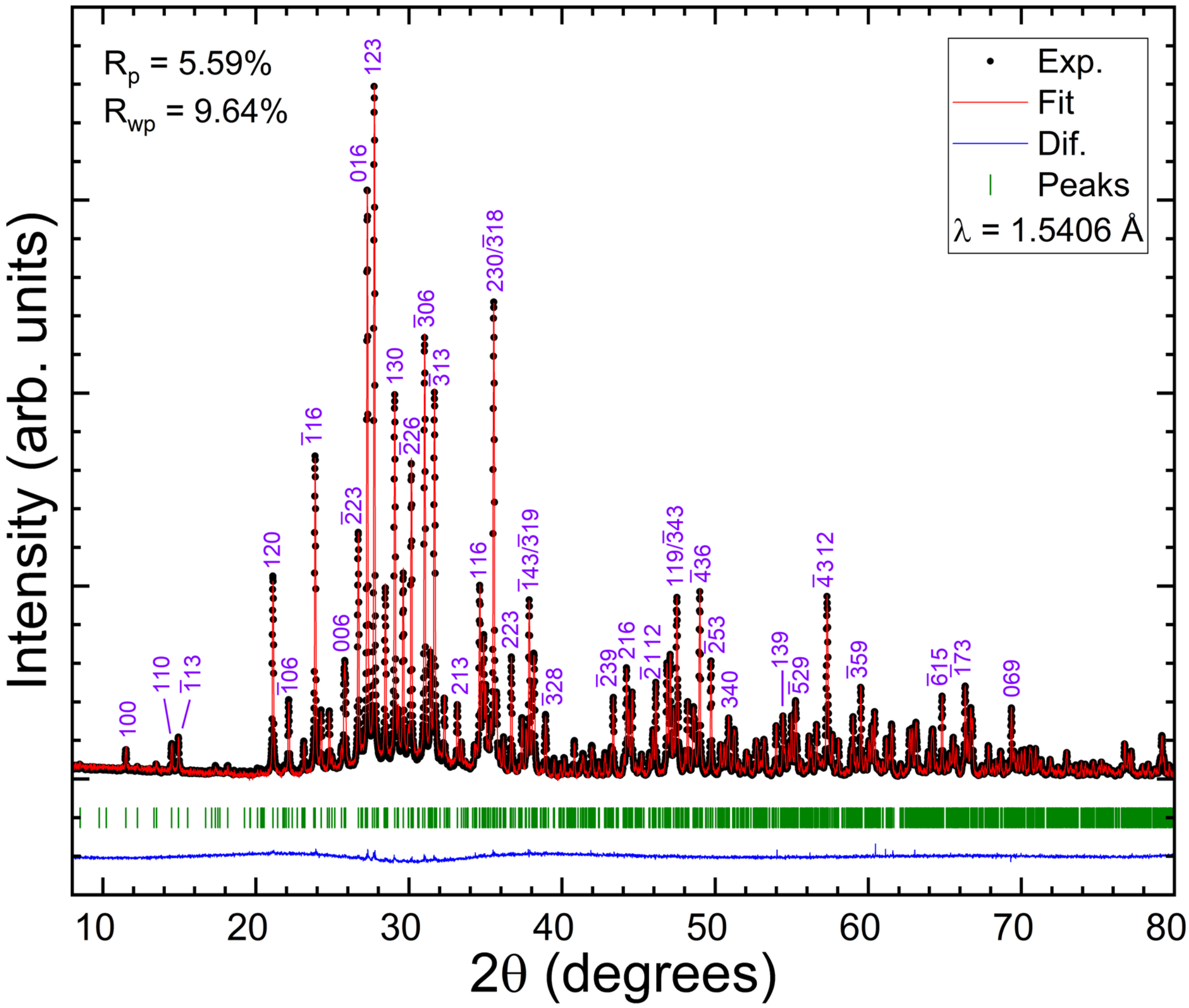

**Figure 1:** XRD pattern measured at ambient conditions. The dots represent the experiment. The red line is the Rietveld refinement, the blue line the residual, and the green ticks identify the Bragg peak positions. R-values of the refinement are given in the plot. The refinement was performed assuming the atomic positions reported in the literature.[20] Selected high-intensity Bragg reflections are labeled with their Miller indices.

The crystal structure of the β' phase is represented in Figure 2. The asymmetric unit contains nine crystallographically independent Mn sites, six independent P sites, and twenty-four independent O sites, all occupying general 4e Wyckoff positions in the monoclinic space group $P2_1/c$.[20] The crystal structure is characterized by a complex three-dimensional framework composed of several distinct manganese coordination polyhedra.[20] Considering Mn-O distances smaller than 2.75 Å, following the structural criterion established by Stephens and Calvo, five different irregular $MnO_6$ octahedra and four different irregular $MnO_5$ polyhedra are identified.[20] The structure contains dimers of edge-sharing $MnO_6$, dimers of edge-sharing Mn-centered $MnO_5$ five-vertex polyhedra, and $MnO_5$–$MnO_6$–$MnO_5$ trimers extending parallel to the crystallographic *c*-axis. The phosphate groups are present as isolated $PO_4$ tetrahedra that connect to the manganese-oxygen framework exclusively through shared oxygen vertices, thereby contributing to the structural stability of the network. The average P–O bond distance is 1.54 Å, and the value for each individual bond within the six $PO_4$ tetrahedra differs by no more than 0.01 Å.

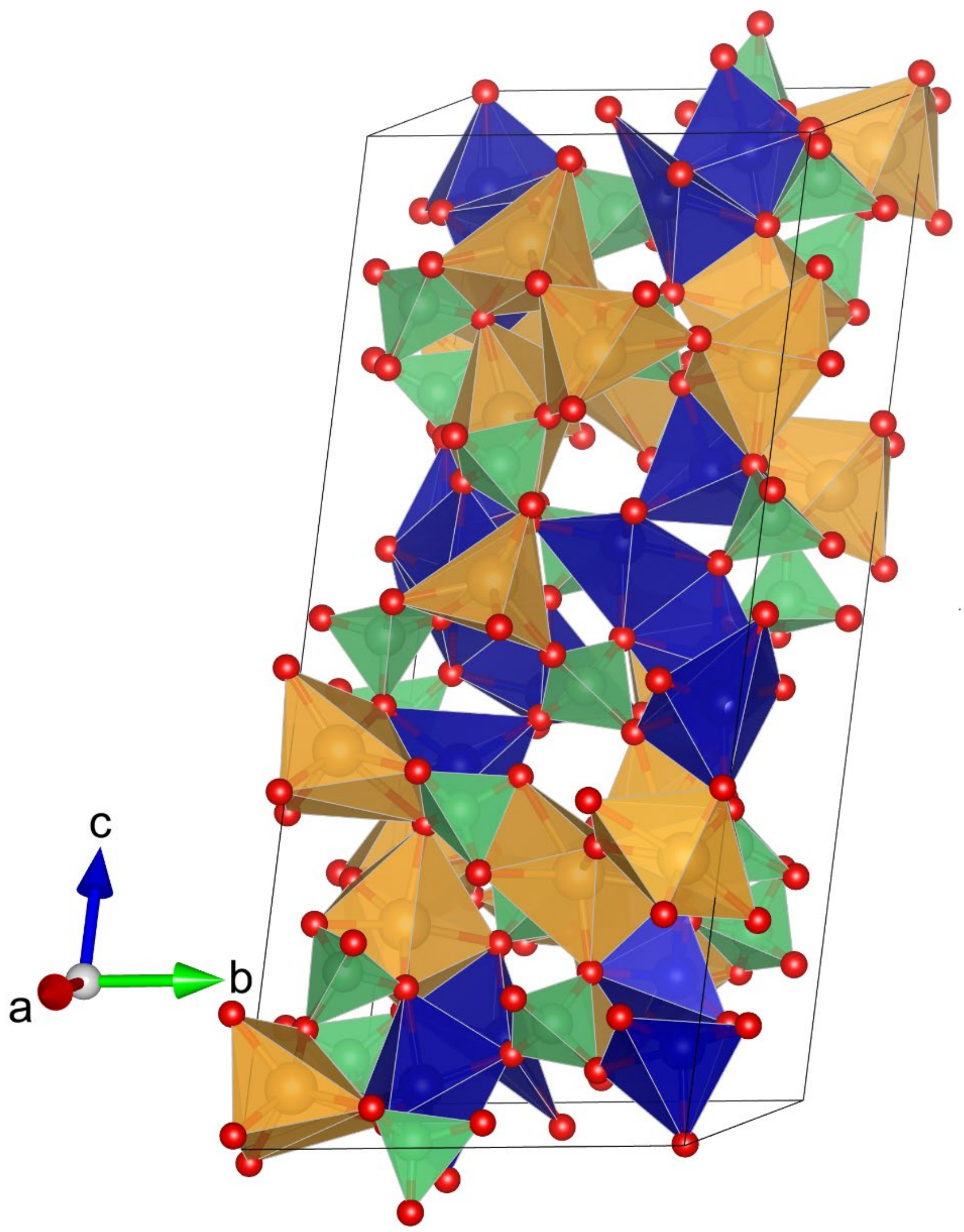

**Figure 2:** Crystal structure of β'-$Mn_3(PO_4)_2$. $MnO_6$ octahedra are represented in yellow and $MnO_5$ polyhedra in blue. The $PO_4$ tetrahedra are shown in green. Red spheres represent oxygen atoms.

**c. High-pressure powder XRD**

A membrane-type diamond-anvil cell (DAC) with 500 μm culet diamonds was employed for compression experiments at room temperature. Stainless steel was used as the gasket material, pre-indented to a thickness of 50 μm and drilled to create a high-pressure chamber with a diameter of 200 μm for the sample. The sample was placed at the center of this chamber along with a copper (Cu) grain serving as a pressure standard. A 4:1 methanol–ethanol mixture was used as the pressure-transmitting medium.[21] Experiments were conducted at the Extreme Conditions beamline ID15B of the European Synchrotron Radiation Facility, using a monochromatic X-ray wavelength of 0.4097 Å. XRD data were recorded with an EIGER2 X 9M CdTe flat-panel detector (340 × 370 mm). The sample-to-detector distance was calibrated using XRD data from a silicon (Si) standard, following standard procedures within the DIOPTAS software suite. At each pressure point, two XRD patterns were collected: one covering the entire sample and another from the Cu standard. The pressure inside the sample chamber was determined from the volumetric compression of the Cu standard.[22]

**d. DFT calculations**

All total energy calculations and geometry optimizations for the antiferromagnetic $Mn_3(PO_4)_2$ system were carried out using spin-polarized density-functional theory (DFT) within the Vienna ab initio Simulation Package (VASP).[23] The projector augmented-wave (PAW) method[24] was employed together with a plane-wave basis set with a kinetic-energy cutoff of 560 eV, ensuring well-converged total energies. The generalized-gradient approximation in the Perdew–Burke–Ernzerhof for solids (PBEsol) formulation[25] was adopted for the exchange-correlation functional. The Brillouin zone was sampled using a 4 × 2 × 2 Monkhorst–Pack k-point mesh.[26] The electronic self-consistent cycle was converged to within $10^{-6}$ eV, and structural relaxations were continued until the residual Hellmann–Feynman forces on all atoms were smaller than 0.005 eV $Å^{-1}$. To account for the strong on-site Coulomb interaction of the Mn 3d electrons, the simplified rotationally invariant GGA+U approach proposed by Dudarev *et al.*[27] was employed with an effective parameter of Ueff = 4.3 eV, a standard value that has been widely adopted in the literature. A sensitivity analysis showed that varying Ueff from 2 to 8 eV changes the calculated bulk modulus by less than 5%, indicating that the results are only weakly dependent on the choice of Ueff.

For each selected volume, corresponding to a given hydrostatic pressure, a full structural optimization was performed. Owing to the complexity of the β'-$Mn_3(PO_4)_2$ unit cell (Z = 12), which contains 156 atoms (nine crystallographically independent Mn atoms, six independent P atoms, and twenty-four independent O atoms occupying 4e Wyckoff positions), all internal atomic coordinates were optimized simultaneously together with the lattice parameters and the monoclinic β angle. The relaxation was performed until the stress tensor became diagonal with equal normal components, ensuring hydrostatic conditions. The optimized structure at each volume was then used to determine the corresponding pressure from the calculated stress tensor and to evaluate the structural and elastic properties.

All calculations were performed at 0 K, as is standard in static DFT simulations. Explicit inclusion of finite-temperature effects would require the calculation of the phonon density of states using large supercells (e.g., 2 × 2 × 2 supercells containing several hundred atoms), followed by quasi-harmonic approximation calculations to evaluate the vibrational free energy and derive thermodynamic properties such as the thermal expansion coefficient. Owing to the substantial computational cost of such calculations, finite-temperature effects were not considered in the present work. This approximation is generally well justified for high-pressure studies of phosphate compounds, for which static 0 K DFT calculations have been shown to provide an accurate description of the structural and compressional behavior measured experimentally at room temperature[15].

The non-magnetic, the ferromagnetic, and several antiferromagnetic spin configurations involving the nine crystallographically independent Mn atoms were examined. Among the configurations considered, the lowest-energy solution corresponded to a ++−− spin arrangement, which was adopted throughout this work. The elastic properties and single-crystal elastic constants were calculated using the method developed by Le Page[28] as implemented in VASP. The resulting second-order elastic stiffness tensor $C_{ij}$ was used to evaluate the Born mechanical stability criteria[29] at ambient pressure. For the monoclinic system considered here, the full expressions of the stability criteria were taken from literature[29]. Under finite hydrostatic pressure, the generalized Born stability criteria[29] were applied. The effect of pressure was incorporated into the elastic constants using the standard transformations for diagonal and off-diagonal tensor components ($C_{ii} \rightarrow C_{ii} - P$ for i = 1, 2...6 and $C_{ij} \rightarrow C_{ij} + P$ for i, j = 1, 2, 3). These modified elastic constants were then used to assess the mechanical stability under compression.

## III. Results

### a. High-pressure structural changes

Figure 3 presents selected XRD patterns collected at different pressures. Over the pressure range from 0.30(3) to 13.02(5) GPa, the diffraction patterns are satisfactorily described using the structural model of the β' phase, refining only the unit-cell and peak-shape parameters. The atomic coordinates were not refined because the diffraction data collected under compression in the diamond-anvil cell did not provide sufficient information to reliably refine the 117 positional parameters (three coordinates for each of the 39 atoms in the unit cell). Instead, the atomic coordinates determined from the ambient-pressure structure[20] were fixed throughout the refinements, while only the unit-cell and peak-shape parameters were allowed to vary.

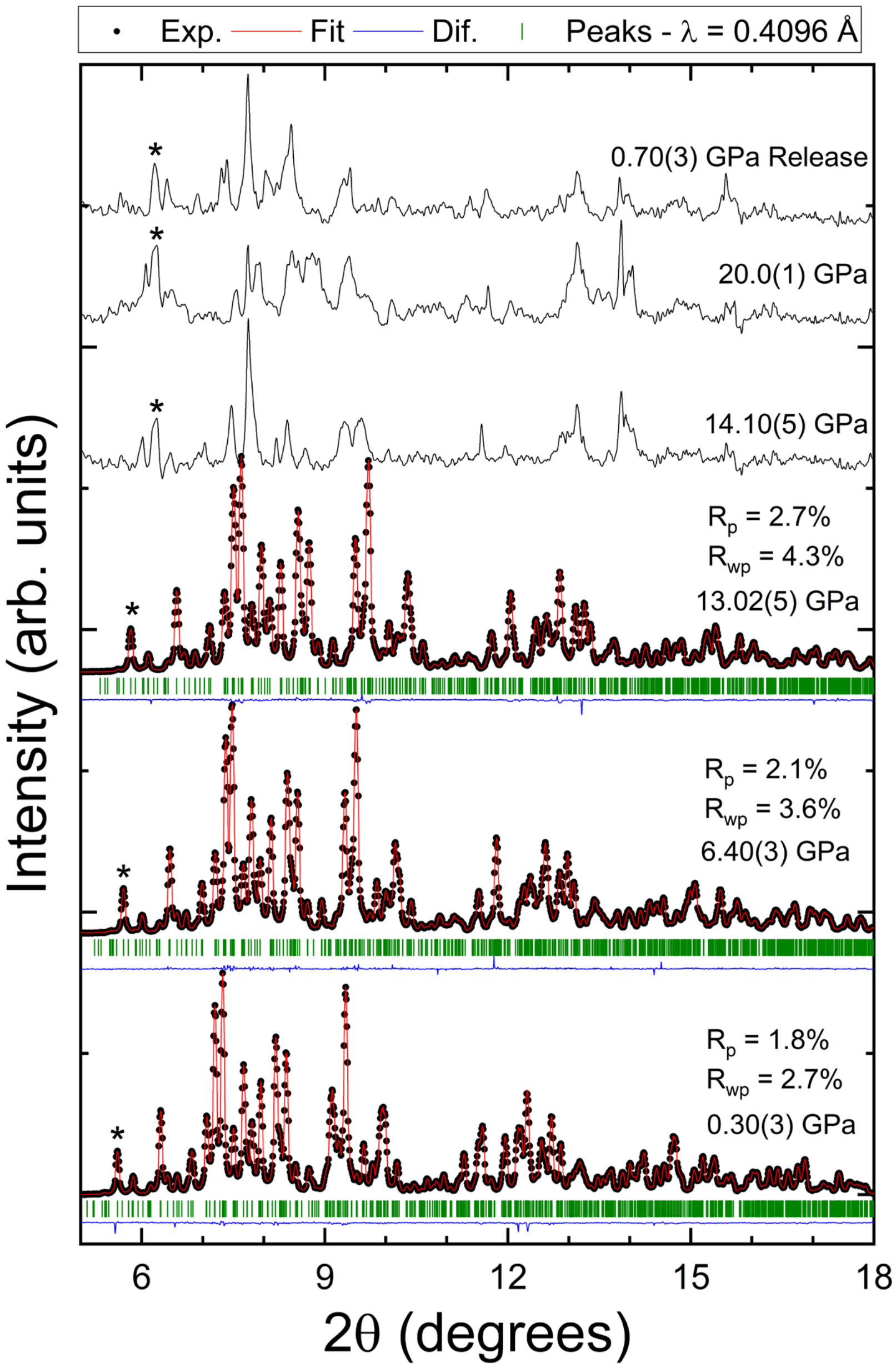

**Figure 3:** XRD patterns measured in situ during compression at 0.30(3), 6.40(3), 13.02(5), 14.10(5), and 20.0(1) GPa. An additional pattern recorded after decompression from 20.0(1) to 0.70(3) GPa is included to demonstrate the irreversible nature of the pressure-induced structural changes. The three lower patterns (0.30(3), 6.40(3), and 13.02(5) GPa) show the evolution of the β' phase before the onset of structural degradation. Dots represent the experiments, the red lines the Rietveld refinements, the blue lines the difference profiles, and the green tick marks the calculated Bragg reflection positions. The refinement residuals (R-values) are given in the figure. The three upper patterns (14.10(5), 20.0(1), and 0.70(3) GPa after decompression), shown as black lines, illustrate the pressure-induced degradation of the XRD pattern, which persists upon further compression and after pressure release. The asterisks denote the reflections used to illustrate peak broadening in Figure 4.

Upon increasing the pressure from 13.02(5) to 14.10(5) GPa, a clear weakening of the XRD signal accompanied by significant peak broadening was observed. In addition, we found that the XRD pattern measured at 14.10(5) GPa cannot be explained assuming the β' phase. These changes in the diffraction patterns upon compression suggest that the sample undergoes a partial loss of long-range crystalline order at 14.10(5) GPa. These observations are consistent with the development of a highly disordered state but do not uniquely distinguish between partial amorphization, nanocrystallization, severe microstrain, defect accumulation, or unresolved phase coexistence. With further compression up to 20.0(1) GPa, the diffraction peaks progressively broaden and decrease in intensity, indicating the development of substantial structural disorder within the material. In the pressure range between 14.10(5) and 20.0(1) GPa, the XRD patterns can no longer be satisfactorily indexed using the crystal structure of β'-$Mn_3(PO_4)_2$ or any other known polymorph of $Mn_3(PO_4)_2$ ($\alpha$, $\gamma$, and $\delta$).[18,19] Likewise, attempts to index the patterns using reported crystal structures of related $A_3(PO_4)_2$-type compounds available in the ICSD database, such as $Cr_3(PO_4)_2$, $Co_3(PO_4)_2$, or $Zn_3(PO_4)_2$, were unsuccessful. A possible decomposition into 3 MnO + $P_2O_5$ was also considered; however, the observed diffraction features are inconsistent with the patterns expected for these phases.

Importantly, the pressure-induced changes observed in the XRD patterns above 14.10(5) GPa are not reversible upon decompression. As shown in Figure 3, a comparison of the XRD patterns collected in situ at 0.30(3) GPa before compression and at 0.70(3) GPa after decompression demonstrates that the structural changes persist after the pressure is released. The recovered sample exhibits broadened and weakened diffraction peaks that differ substantially from those of the starting material, indicating that the HP treatment induces permanent structural modifications. These observations might suggest that compression leads to a metastable disordered or partially amorphous state rather than a well-defined crystalline HP polymorph or simple decomposition products. Although this behavior is consistent with pressure-induced structural disorder, the available diffraction data do not allow us to uniquely determine the microscopic nature of the transformed state.

The possible influence of non-hydrostatic stresses on the pressure-induced changes observed in the XRD patterns above 14.10(5) GPa should be considered. We attribute the observed structural disorder to an intrinsic response of the crystal to compression rather than to the onset of non-hydrostatic conditions. Although the 4:1 methanol–ethanol pressure-transmitting medium gradually loses hydrostaticity above approximately 10 GPa,[21] no significant changes in the diffraction peak widths are observed at this pressure. If non-hydrostatic stresses were the primary cause of the observed disorder, peak broadening would be expected to begin near the hydrostatic limit of the pressure medium.[30] Instead, the XRD pattern collected at 13.02(5) GPa exhibits peak widths comparable to those measured below 10 GPa (Figure 3).

This behavior is quantified in Figure 4, which shows the full width at half maximum (FWHM) of the reflections marked with asterisks in Figure 3. The FWHM remains essentially constant up to 13.02(5) GPa and increases only at 14.10(5) GPa, from 0.053(1)° to 0.065(1)°. Upon further compression, the peaks continue to broaden progressively. Moreover, the increased FWHM is retained after decompression, supporting that the pressure-induced structural changes are irreversible. If peak broadening originated primarily from deviatoric stresses associated with the pressure-transmitting medium, a substantial recovery of the peak widths upon pressure release would be expected.

Further support for this interpretation is provided by previous high-pressure studies of related phosphates. Similar structural changes have been reported below 20 GPa in experiments performed using both 4:1 methanol–ethanol and neon as pressure-transmitting media.[31] Since neon remains quasi-hydrostatic to approximately 20 GPa,[21] the consistency of the reported behavior across different pressure media indicates that the observed structural changes are intrinsic to compression rather than artifacts arising from non-hydrostatic stresses.

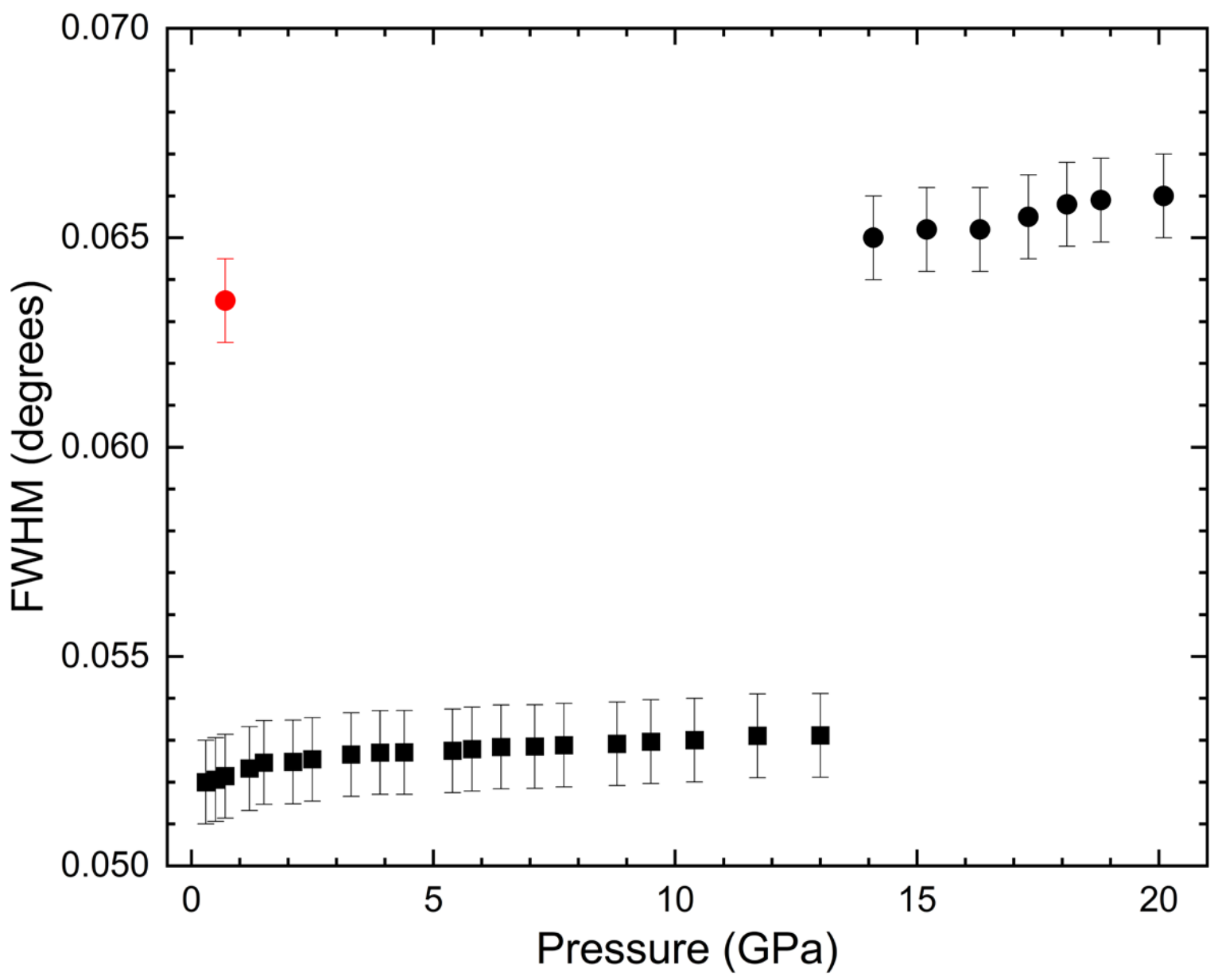

**Figure 4:** Full width at half maximum (FWHM) of XRD peaks identified in Figure 3 by asterisks as a function of pressure. Data up to 13.02(5) GPa are represented by squares. Data from 14.10(5) to 20.0(1) GPa are represented by circles. The red circle was obtained from the experiment performed at 0.70(3) GPa after decompression.

The pressure-induced disorder observed in the crystal structure of β'-$Mn_3(PO_4)_2$ at 14.10(5) GPa closely resembles the loss of long-range crystalline order reported for β-$Zn_3(PO_4)_2$ near 14 GPa.[16] In contrast, this behavior differs markedly from that of $Co_3(PO_4)_2$, which undergoes a first-order transition to another crystalline phase under compression.[15] The different HP behaviors of β'-$Mn_3(PO_4)_2$, β-$Zn_3(PO_4)_2$, and $Co_3(PO_4)_2$ can be rationalized in terms of their structural similarities and differences. In particular, the observation that the Mn and Zn compounds develop structural disorder at comparable pressures, whereas the Co compound undergoes a first-order crystalline-to-crystalline phase transition, appears to be closely related to the topology and flexibility of their crystal frameworks. β'-$Mn_3(PO_4)_2$ and β-$Zn_3(PO_4)_2$ share similarly complex monoclinic frameworks composed of distorted $AO_5$/$AO_6$ polyhedra connected to isolated $PO_4$ tetrahedra. Both structures possess relatively large unit cells, multiple crystallographically distinct metal sites, and significant polyhedral distortions. These characteristics produce structurally flexible frameworks that can accommodate compression through progressive local distortions rather than through a cooperative reconstructive transition into another long-range ordered phase. As pressure increases, accumulated strain and distortion likely destabilize the long-range periodicity of the lattice, leading to pressure-induced irreversible structural disorder, possibly involving partial amorphization, at similar pressures in both compounds. This behavior is further supported by the analogous response observed in $Mn_3(VO_4)_2$,[12] which also undergoes pressure-induced amorphization, in contrast to other $A_3(VO_4)_2$ vanadates that typically do not exhibit amorphization under compression.[13,32] This parallel suggests that the tendency toward pressure-induced disorder may be an intrinsic feature of structurally complex Mn-based frameworks.

In contrast, $Co_3(PO_4)_2$ crystallizes in a simpler and more ordered $\gamma$-$Zn_3(PO_4)_2$-type framework with a smaller unit cell and fewer crystallographic distinct cation sites. This simpler topology facilitates a cooperative structural rearrangement under compression, enabling the system to transform into another crystalline phase through a first-order phase transition rather than losing long-range order. The comparatively regular $CoO_5$/$CoO_6$ coordination environments may also favor a more collective and symmetry-driven transformation mechanism. Therefore, the distinct pressure responses suggest that structural complexity and framework flexibility play a key role in determining the compression mechanism. Complex and highly distorted frameworks, such as those of the Mn and Zn $\beta$-phases, appear more prone to pressure-induced disorder, whereas simpler and more ordered frameworks, such as that of $Co_3(PO_4)_2$, can more readily accommodate compression through crystalline-to-crystalline transitions.

The comparison among $\beta'$-$Mn_3(PO_4)_2$, $\beta$-$Zn_3(PO_4)_2$, and $Co_3(PO_4)_2$ suggests that framework topology may play an important role in determining the high-pressure response of $A_3(PO_4)_2$ phosphates. The Mn and Zn compounds possess comparatively large unit cells, multiple crystallographically distinct cation sites, and highly distorted coordination polyhedra, whereas $Co_3(PO_4)_2$ adopts a structurally simpler framework. Although this comparison does not establish a universal mechanism, it suggests that structurally complex frameworks may be less able to accommodate compression through cooperative crystalline-to-crystalline phase transitions and instead evolve toward states characterized by a degradation of long-range crystallographic order. Additional high-pressure studies on related phosphates will be required to assess the generality of this trend.

**b. Compressibility and Structural Evolution Under Pressure**

From XRD patterns, we obtained the change of unit-cell parameters with pressure. The results are represented in Figure 5. In the figure we compare experiments (symbols) with DFT calculations (lines), which give very similar results. Fig. 5(a) shows that compression is anisotropic. Fig. 5(b) shows that be $\beta$ angle has a non-linear behavior but stays close to 120.8º. The pressure dependence of the volume shown in Fig. 5(c) was described by a third-order Birch-Murnaghan (BM3) equation of state (EoS).[33] The values obtained for the volume at 0 GPa, $V_0$, the bulk modulus at 0 GPa, $K_0$, and its pressure derivative, $K_0$', are given in Fig. 5(c). From experiments we obtained $V_0$ = 1861(1) Å$^3$, $K_0$, = 81(2) GPa, and $K_0$' = 3.3(2). From calculations we obtained $V_0$ = 1869(1) Å$^3$, $K_0$ = 86(3) GPa, and $K_0$' = 2.9(4). The 3$\sigma$ confidence ellipses of $K_0$ and $K_0$' obtained from the BM3 EoS fits, show significant overlap (see the inset of Fig. 5(c)). This substantial overlap indicates excellent agreement between the two datasets, with the fitted parameters being statistically indistinguishable within their uncertainties. Consequently, the BM3 fits provide consistent estimates of the elastic properties across the two approaches.

The bulk modulus determined from experiments for $Mn_3(PO_4)_2$, 81(2) GPa, is very similar to that of $Co_3(PO_4)_2$, 78(2) GPa,[15] and $Ca_3(PO_4)_2$, 79.5(2) GPa.[35] When compared with the corresponding vanadates, the phosphates exhibit lower bulk moduli: $Ca_3(VO_4)_2$, 99(2) GPa,[36] $Mn_3(VO_4)_2$, 101(1) GPa,[12] and $Co_3(VO_4)_2$, 122(3) GPa.[25] A similar trend is observed for zircon-type compounds, where $VO_4$-based vanadates are consistently stiffer than $PO_4$-based phosphates.[9,10] The difference is systematic and can be attributed to the slightly shorter and stronger V–O bonds in $VO_4$ tetrahedra compared to P–O bonds in $PO_4$, which makes vanadate tetrahedra more resistant to compression. In addition, vanadates have slightly smaller unit-cell volumes, further enhancing the lattice stiffness.

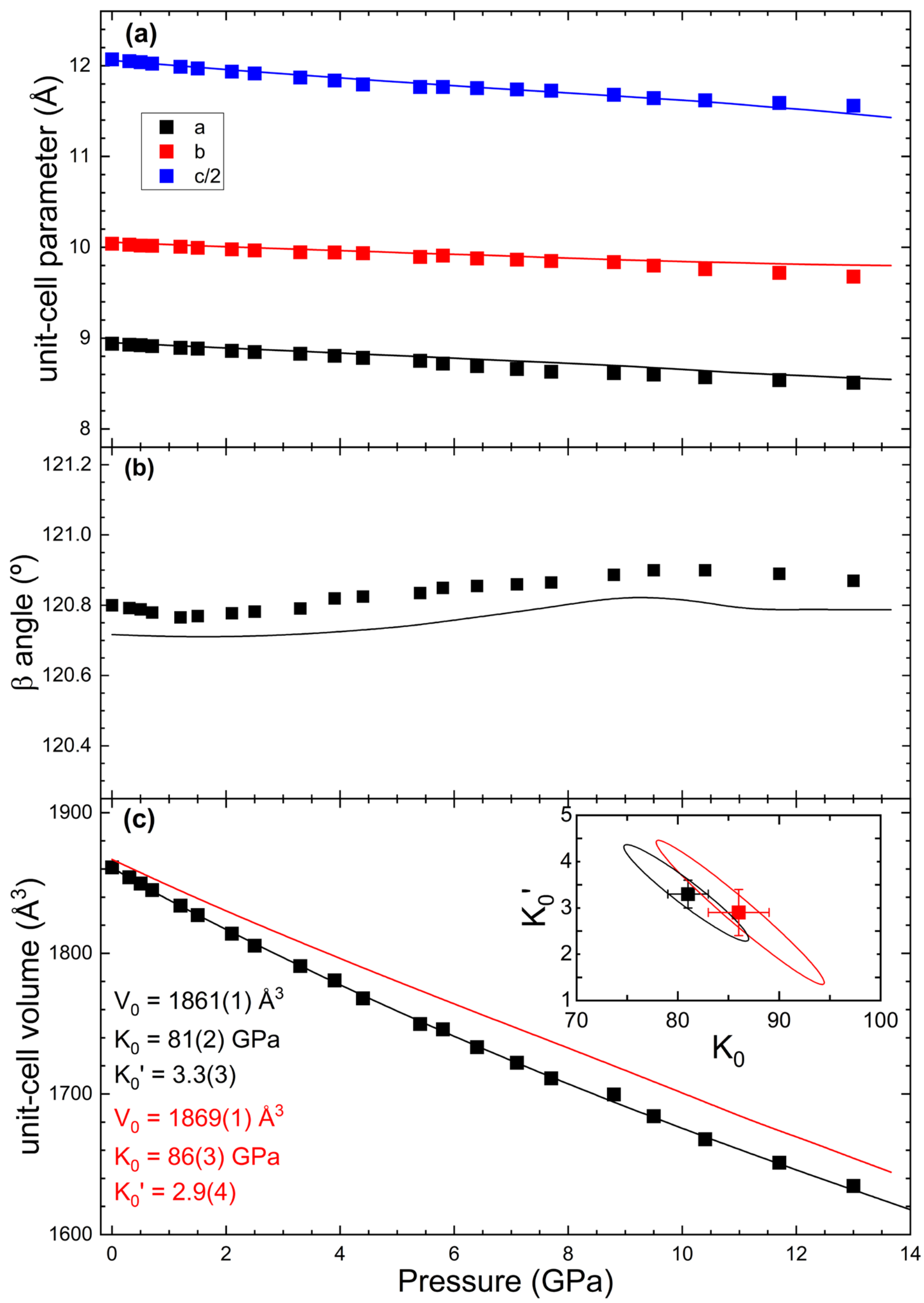

**Figure 5:** (a) Pressure dependence of unit-cell parameters *a*, *b*, and *c*. For the *c*-axis, we represented c/2 to facilitate comparison. (b) Pressure dependence of the $\beta$ angle. (c) Pressure dependence of the unit-cell volume. Symbols are results from experiments and lines results from calculations. For the volume, the red line shows results from calculations and the black line the Birch-Murnaghan fit to experiments. The inset plots $K_0'$ versus $K_0$. It includes the 3-sigma confidence ellipse from fits to experiments and calculations. The EoS parameters determined from experiments and calculations are included in the figure in black and red, respectively.

Since the crystal structure of β'-$Mn_3(PO_4)_2$ is monoclinic, its compressibility must be described using the compressibility tensor, which contains four independent components.[37] The eigenvectors of this tensor define the principal axes of compressibility, while the corresponding eigenvalues provide the linear compressibilities along these directions. Using our experimental data and the PASCAL software,[38] we determined these principal axes and their linear compressibilities, which are summarized in Table 1. Two of the axes ($X_1$ and $X_2$) lie in the $\overline{ac}$ plane. $X_2$ is the most compressible axis. It makes an 8º angle with the *a*-axis and has a compressibility 80% larger than the least compressible axis, the *b*-axis. X1 makes a 53º angle with the *a*-axis.

| $\kappa_1$ = 3.5(2) $10^{-3}$ GPa | $X_1$ = [3 0 4] |
|---|---|
| $\kappa_2$ = 4.3(2) $10^{-3}$ GPa | $X_2$ = [10 0 1] |
| $\kappa_3$ = 2.4(1) $10^{-3}$ GPa | $X_3$ = [0 1 0] |

**Table 1:** Main axes of compressibility ($X_i$) and their compressibility ($\kappa_i$).

In β'-$Mn_3(PO_4)_2$, the anisotropic compressibility comes from how the rigid $PO_4$ tetrahedra and the more deformable Mn–O polyhedra are connected in different crystallographic directions. The key point is that $PO_4$ tetrahedra are very rigid because P–O bonds are short and strongly covalent. Under pressure, phosphate tetrahedra compress very little compared with Mn–O bonds (see next subsection). Along the *b*-direction, the structure contains a more continuous and strongly connected polyhedral framework (see Figure 2). In β'-$Mn_3(PO_4)_2$, chains/networks of edge- and corner-sharing $MnO_5$/$MnO_6$ polyhedra are reinforced by $PO_4$ tetrahedra along certain directions. Since these rigid tetrahedra are densely linked along *b*, compression along this axis requires shortening relatively incompressible P–O-supported connections. Compression in the perpendicular main axes can occur by polyhedral tilting/hinging. These mechanisms reduce the need for direct bond compression, making those axes more compressible.

### c. Local Structural Changes Under Pressure

To better understand the pressure-induced changes in the crystal structure, it is essential to analyze the evolution of the bond distances under compression. This information cannot be reliably extracted from the powder XRD experiments as described in subsection 3.a. Nevertheless, the excellent agreement between the experimental results and the calculated pressure dependence of the unit-cell parameters supports the use of DFT-derived bond distances to investigate their pressure evolution. As shown in Figure 6, the simulated bond lengths at ambient pressure are consistent with those obtained from accurate XRD results reported in the literature.[20] At ambient pressure, most of the Mn-O bond distances range between 2.03 Å and 2.49 Å. According to Stephens and Calvo[20], which include distances up to 2.75 Å in the first coordination shell, there are five Mn atoms that can be considered in octahedral coordination. These are the atoms identified as Mn1A, Mn1B, Mn1C, Mn2B, and Mn3B by them and in Figure 6. The other four Mn atoms, Mn2A, Mn2C, Mn3A, and Mn3C, are penta-coordinated with a sixth oxygen atom at a distance from the Mn atom ranging between 2.78 and 3.02 Å. The nomenclature of the crystallographically distinct Mn sites (Mn1A–Mn3C) follows that introduced by Stephens and Calvo[20] and is adopted throughout this work.

Figure 6 shows that the structural response to compression involves substantial modifications of several Mn coordination polyhedra rather than being localized at a single crystallographic site. With increasing pressure, the $MnO_6$ octahedra around Mn1B and Mn1C become progressively more regular, approaching the nearly symmetric octahedral geometry already exhibited by Mn1A, Mn2B, and Mn3B at ambient pressure. In contrast, three initially penta-coordinated Mn sites (Mn2A, Mn2C, and Mn3A) undergo a gradual increase in coordination as oxygen atoms initially located at distances greater than 2.75 Å move closer to the Mn centers with increasing pressure. Consequently, these sites attain an effective octahedral coordination at approximately 4.7, 7.6, and 7.4 GPa, respectively. By 14 GPa, immediately before the onset of the experimentally observed degradation of long-range crystallographic order, the pressure-induced $MnO_6$ octahedra closely resemble the other five octahedra present in the structure. In contrast, the ninth Mn site, Mn3C, remains penta-coordinated throughout the pressure range investigated. These results indicate that the structural response is distributed throughout several Mn coordination environments and reflects a collective rearrangement of the framework rather than an isolated distortion of a single Mn site. The progressive increase in the coordination number of three Mn atoms demonstrates the ability of the framework to accommodate compression through local polyhedral rearrangements prior to the experimentally observed degradation of long-range crystallographic order.

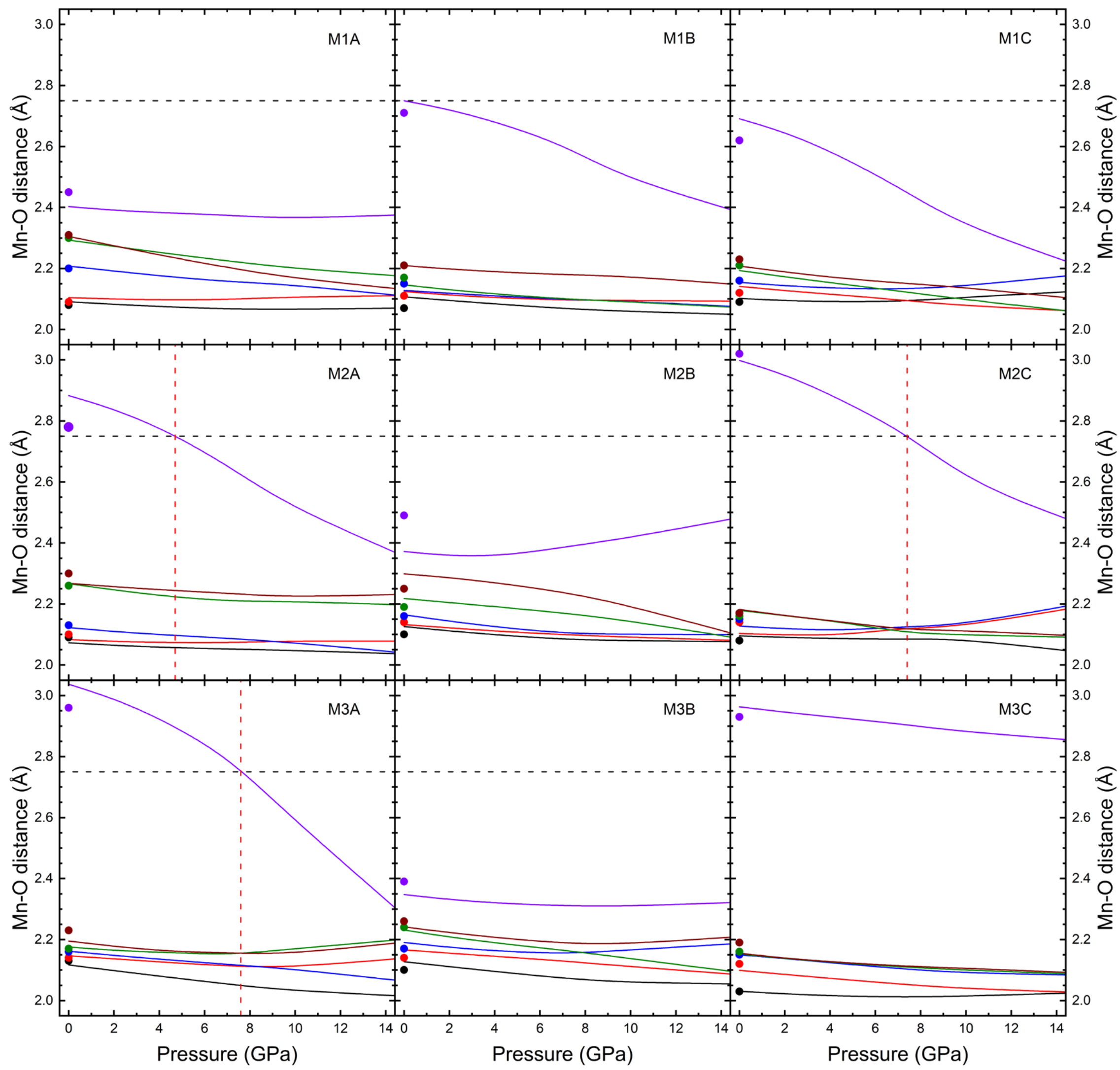

**Figure 6:** Pressure dependence of Mn-O bond distances for the nine independent Mn atoms. Symbols are results taken from ambient pressure XRD experiments.[20] The lines are the results of our calculations. Mn atoms are identified using the nomenclature proposed by Stephens and Calvo.[20] Different colors identify the Mn–O bond distances for the distinct Mn coordination polyhedra and are used only to facilitate visual comparison. The horizontal black dashed lines mark the distance used in the literature as a limit for the first coordination shell. The vertical dashed red lines denote the pressures where penta-coordinated Mn atoms become octahedrally coordinated.

Another fact to highlight is the pressure evolution of the $MnO_6$ octahedra of atom Mn2B. This polyhedron exhibits an unusual mechanism in which two Mn–O bonds contract substantially, while the longest Mn–O bond anomalously expands. Such behavior is indicative of anisotropic bond softening and suggests that the octahedra accommodate compression through distortion rather than rigid contraction. The elongation of the weakest Mn–O linkage likely reflects a reduction in bond stiffness and the development of a soft local structural mode. As pressure increases, these enhanced local distortions might disrupt the cooperative connectivity between neighboring octahedra, which might promote structural disorder and contribute to a loss of long-range crystallographic order such as that observed in our XRD experiments.

We have also analyzed the pressure dependence of the volume of Mn coordination polyhedra, considering all of them as octahedra, and the results are summarized in Figure 7. The figure also includes the pressure evolution of the $PO_4$ tetrahedral volume, which is representative of all six tetrahedra, as they exhibit similar pressure-dependent behavior. In the figure, we show that the $MnO_6$ units are much more compressible than the $PO_4$ tetrahedra, which behave as rigid incompressible units as in most phosphates.[39] In most of the $MnO_6$ octahedra, the polyhedral volume decreases between 8% and 14% from 0 to 14 GPa. However, there are three polyhedra showing an unusual large compressibility and an unusual decrease of compressibility as pressure increases. These are the polyhedral units associated with Mn2A, Mn2B, and Mn3A atoms. In particular, the polyhedron of Mn2B shows a sharp increase of compressibility around 10 GPa, which we believe might be related to structural changes found at 14.1 GPa. In fact, this behavior might indicate the onset of pressure-induced elastic softening associated with weakening of the bonding network under compression. The reduction in resistance to volume change suggests the development of low-energy deformation mechanisms within the lattice, such as bond-angle distortions and local structural instabilities, which enhance compressibility prior to the loss of crystallinity. Therefore, the observed decrease in bulk modulus might indicate the development of elastic softening preceding structural disorder.

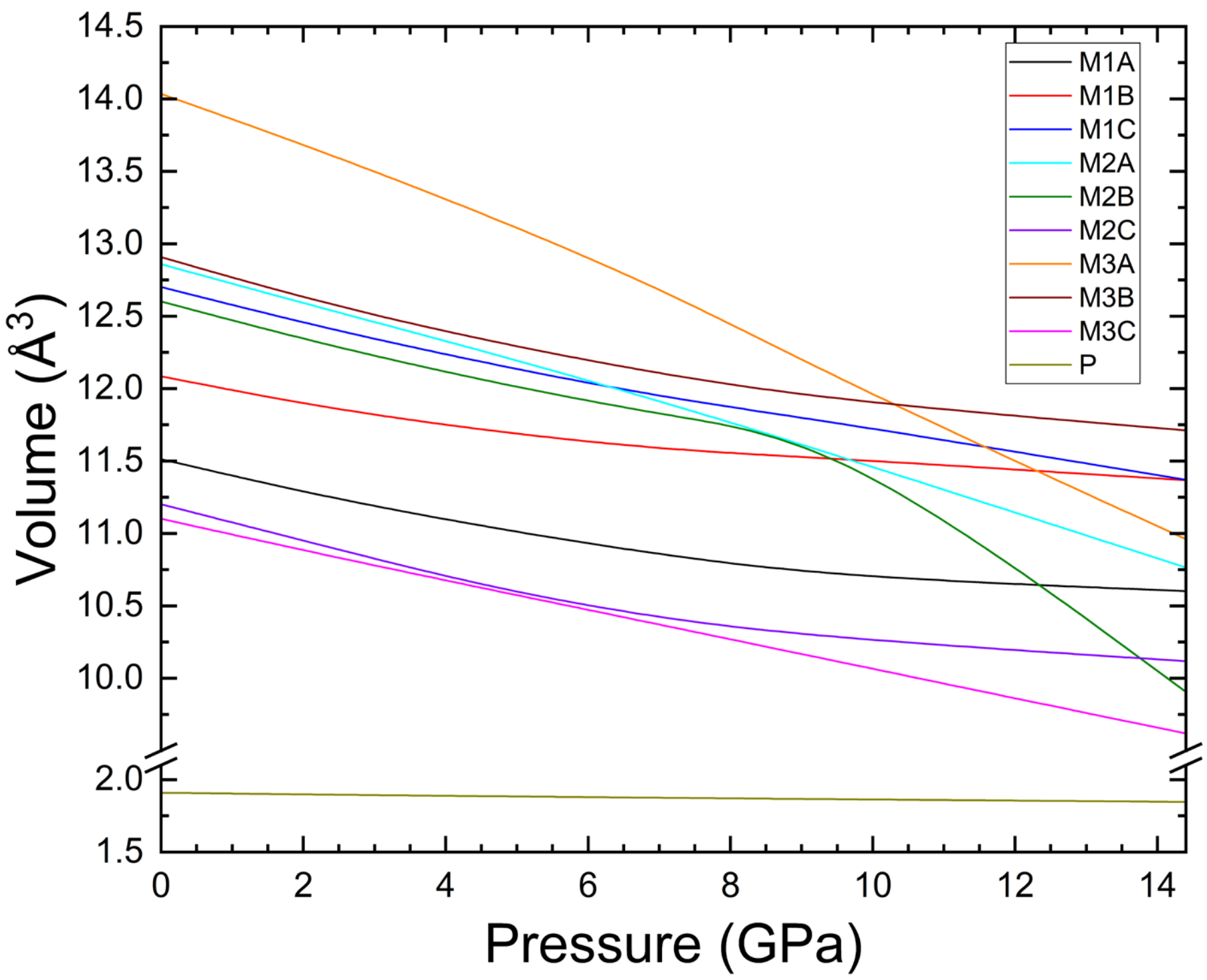

**Figure 7:** Pressure dependence of polyhedral units considering all Mn atoms as octahedrally coordinated.

To close this section, we would like to note that the distinct compressibilities of the $MnO_6$ and $PO_4$ polyhedral units play a key role in determining the compressional behavior of β'-$Mn_3(PO_4)_2$. The rigid $PO_4$ tetrahedra act as nearly incompressible units, while the softer $MnO_6$ polyhedra accommodate most of the pressure-induced volume reduction. This polyhedral contrast controls both the bulk modulus and the anisotropic axial compressibility, with the largest axial contraction occurring along directions dominated by $MnO_6$ distortions.

**d. Pressure-induced evolution of the electronic structure and chemical bonding**

To obtain further insight into the microscopic origin of this structural response, we investigated the evolution of the electronic structure and chemical bonding under compression using the electron localization function (ELF) and spin-polarized projected density of states (PDOS), as shown in Figures 8 and 9. Together, these analyses provide a comprehensive understanding of the evolution of chemical bonding and electronic structure under hydrostatic compression.

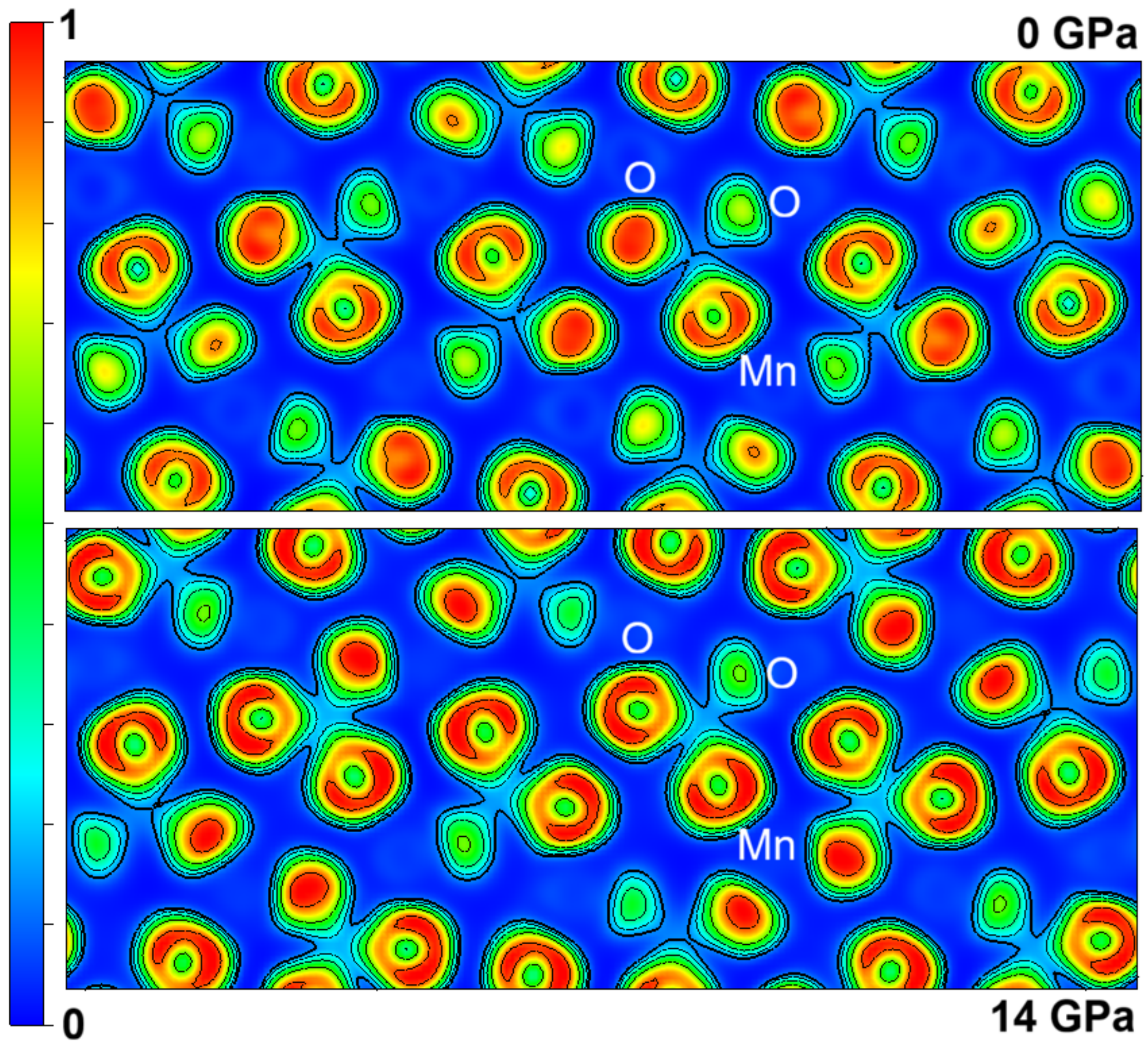

**Figure 8:** Electron localization function (ELF) maps of β'-$Mn_3(PO_4)_2$ at 0 GPa (top) and 14 GPa (bottom). The color scale ranges from 0 (blue) to 1 (red). Strong electron localization around the oxygen atoms indicates the highly covalent nature of the P–O bonds, while the lower localization along the Mn–O bonds is consistent with their predominantly ionic character. The ELF distribution changes only slightly under compression, demonstrating the robustness of the local bonding topology.

The ELF maps (Figure 8) show that the highest electron localization (ELF ≈ 1) is concentrated around the oxygen atoms at both pressures, corresponding to the lone-pair electrons and the highly covalent P–O bonds within the phosphate tetrahedra. In contrast, the Mn–O bond regions exhibit comparatively low ELF values and no pronounced localization maxima along the bond paths, indicating that the Mn–O interactions remain predominantly ionic. Comparison of the ELF maps at 0 and 14 GPa reveals only minor changes in the electron localization. The strong localization associated with the phosphate groups is essentially preserved, while the slight compression of the ELF contours reflects the reduced interatomic distances produced by lattice contraction rather than a significant modification of the bonding character. These results indicate that the $PO_4$ tetrahedra remain electronically robust under compression and that the pressure-induced structural response is accommodated primarily by distortions of the Mn coordination polyhedra.

The pressure dependence of the electronic structure is further illustrated by the spin-polarized PDOS (Figure 9). At both pressures, $Mn_3(PO_4)_2$ remains a semiconductor with an indirect band gap that increases slightly from 3.276 eV at ambient pressure to 3.362 eV at 14 GPa. The valence band is dominated by hybridized Mn 3d and O 2p states, whereas the conduction band is composed mainly of Mn 3d states. Hydrostatic compression produces a moderate broadening of both the Mn 3d and O 2p bands, consistent with the enhanced orbital overlap expected from the shortened interatomic distances. However, the orbital character of the electronic states remains essentially unchanged, indicating that compression modifies the bandwidths rather than the nature of the chemical bonding. The persistence of spin asymmetry in the Mn 3d states further indicates that the localized magnetic moments are preserved throughout the investigated pressure range.

The combined ELF and PDOS analyses therefore show that hydrostatic compression up to 14 GPa produces only modest changes in the electronic structure while preserving the mixed ionic–covalent bonding character of β'-$Mn_3(PO_4)_2$. The rigidity of the covalently bonded $PO_4$ tetrahedra and the comparatively greater flexibility of the Mn coordination polyhedra are consistent with the structural evolution observed experimentally and in the DFT structural optimizations. These results suggest that the experimentally observed degradation of long-range crystallographic order is associated with the progressive accumulation of structural distortions and the loss of mechanical stability, rather than with a pressure-induced electronic instability.

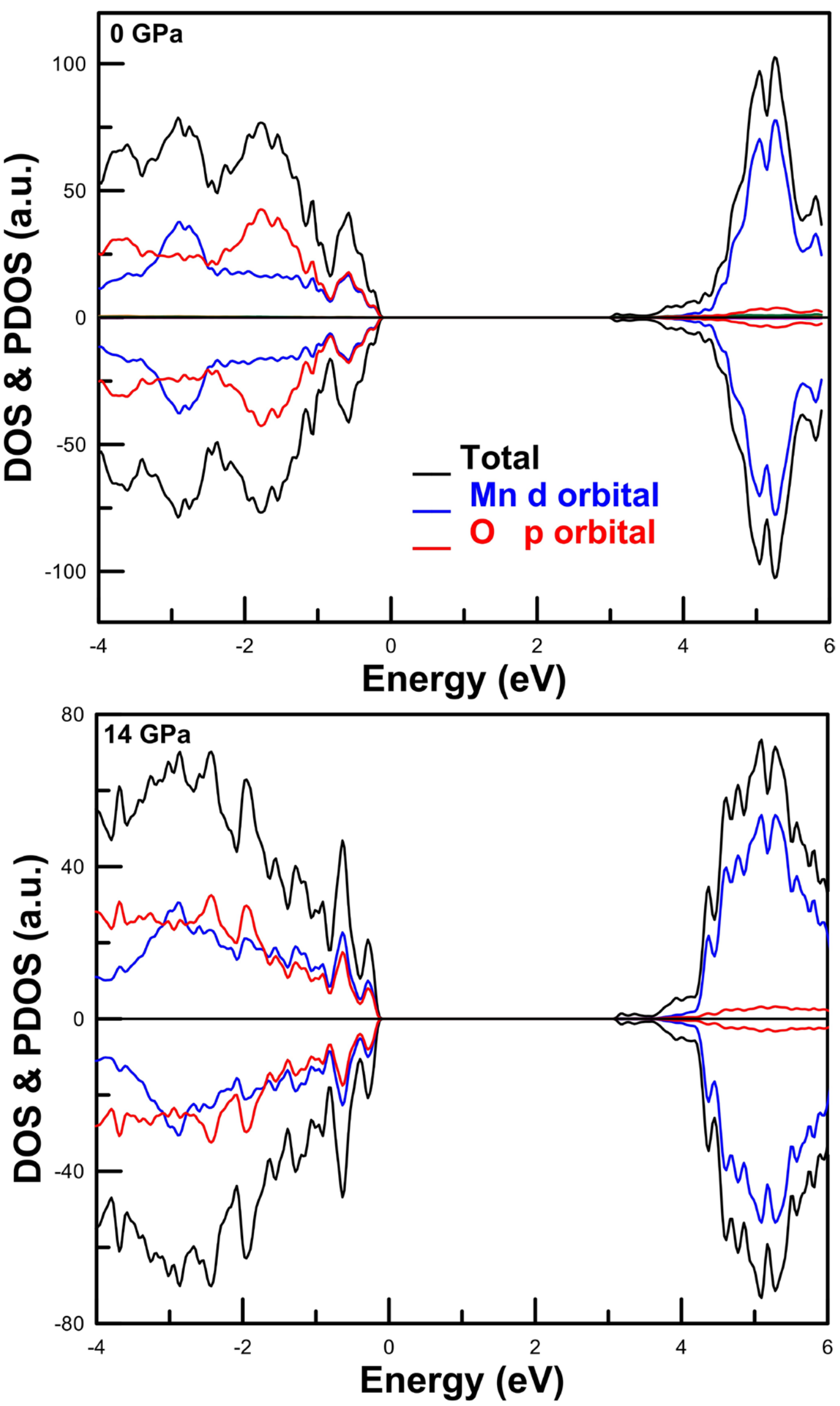

**Figure 9.** Spin-polarized total density of states (DOS, black lines) and projected density of states (PDOS) of the Mn 3d (blue lines) and O 2p (red lines) orbitals of β'-$Mn_3(PO_4)_2$ at 0 GPa (top) and 14 GPa (bottom). Positive and negative values correspond to the spin-up and spin-down channels, respectively, and the Fermi level is set to 0 eV. The valence band is mainly composed of hybridized Mn 3d and O 2p states, while the conduction band is dominated by Mn 3d states. Compression to 14 GPa produces moderate band broadening and a slight increase in the band gap, while preserving the orbital character and spin polarization of the electronic states.

## e. Elastic constants

To assess the stability of β'-$Mn_3(PO_4)_2$ and gain further insight into its anisotropic structural behavior under HP, we calculated the elastic properties of this compound. The results are presented in Tables 2 and 3. At ambient pressure, all eigenvalues of the elasticity tensor are positive. In addition, the Born criteria for monoclinic structures are satisfied.[40] The two results support the mechanical stability of the studied compound. Elastic constants calculations support the mechanical stability up to 13 GPa. We obtained that $C_{33} > C_{11} > C_{22}$, which is consistent with the fact that the *b*-axis is more compressible than the other two main axes of compressibility.

| $C_{11}$ | $C_{12}$ | $C_{13}$ | $C_{15}$ | $C_{22}$ | $C_{23}$ | $C_{25}$ |
|---|---|---|---|---|---|---|
| 151.5 | 73.9 | 64.4 | 10.6 | 159.7 | 79.1 | −2.3 |
| $C_{33}$ | $C_{45}$ | $C_{44}$ | $C_{46}$ | $C_{55}$ | $C_{66}$ | |
| 140.6 | 0.4 | 42.1 | −1.6 | 34.5 | 29.8 | |

**Table 2:** Calculated Elastic Constants at ambient pressure in GPa.

| $C_{11}$ | $C_{12}$ | $C_{13}$ | $C_{15}$ | $C_{22}$ | $C_{23}$ | $C_{25}$ |
|---|---|---|---|---|---|---|
| 138.8 | 107.4 | 73.4 | 30.7 | 164.7 | 97.2 | −12.3 |
| $C_{33}$ | $C_{35}$ | $C_{44}$ | $C_{46}$ | $C_{55}$ | $C_{66}$ | |
| 96.2 | 6.8 | 25.8 | 3.3 | 14.4 | 18.5 | |

**Table 3:** Calculated Elastic Constants at 14.3 GPa in GPa.

The calculated elastic constants at 14.3 GPa indicate that the structure becomes mechanically unstable near this pressure. In particular, one eigenvalue of the elastic tensor becomes negative, revealing the onset of elastic instability. Consistently, one of the generalized Born stability criteria is no longer satisfied, implying that the elastic stiffness matrix loses its positive definiteness. The violated stability criteria for the monoclinic case is

$$2\times[C_{15}C_{25}(C_{33}C_{12} - C_{13}C_{23}) + C_{15}C_{35}(C_{22}C_{13} - C_{12}C_{23}) + C_{25}C_{35}(C_{11}C_{23} - C_{12}C_{13})] - [C_{15}^2(C_{22}C_{33} - C_{23}^2) + C_{25}^2(C_{11}C_{33} - C_{13}^2) + C_{35}^2(C_{11}C_{22} - C_{12}^2)] + C_{55}[C_{11}C_{22}C_{33} - C_{11}C_{23}^2 - C_{22}C_{13}^2 - C_{33}C_{12}^2 + 2C_{12}C_{13}C_{23}] > 0$$

Physically, this means that certain strain modes can lower the elastic energy, rendering the crystal mechanically unstable against small perturbations. The predicted loss of elastic stability occurs at approximately the same pressure at which the diffraction patterns exhibit pronounced peak broadening and intensity loss. This correspondence is consistent with a link between elastic instability and the observed degradation of long-range crystallographic order, although it does not by itself establish the microscopic nature of the pressure-induced transformation.

The macroscopic elastic moduli can be derived from the full set of single-crystal elastic constants, which describe the response of a material to small deformations along different crystallographic directions. Because real materials are often polycrystalline or effectively isotropic at the macroscopic scale, an averaging scheme is required to convert the anisotropic single-crystal elastic tensor into effective isotropic elastic properties. In this work, we employed the Hill approximation,[41] which takes the arithmetic mean of the Voigt (uniform strain) and Reuss (uniform stress) bounds, providing a reliable estimate for polycrystalline aggregates.

Using this approach, we obtained a bulk modulus $B$ = 88.0 GPa, a shear modulus $G$ = 36.3 GPa, a Young's modulus $E$ = 97.0 GPa, and a Poisson's ratio $\nu$ = 0.335. The bulk modulus quantifies the resistance of the material to uniform compression, while the shear modulus reflects its resistance to shape deformation at constant volume. Young's modulus provides a measure of stiffness under uniaxial stress, and Poisson's ratio describes the coupling between axial and transverse strain during deformation.

A notable feature of the results is that $B \gg G$, indicating a strong mechanical asymmetry between volumetric and shear responses. This implies that the material is significantly more resistant to uniform compression than to shear deformation. In practical terms, isotropic compression requires substantially more energy than shear distortion, suggesting that bonding in the system more effectively resists volume reduction than angular or shear distortions of the lattice. Such mechanical imbalance is often relevant to the onset of strain localization and defect formation under external pressure or stress.

In particular, the relatively low shear resistance might indicate a predisposition toward shear instabilities under increasing pressure, which can act as a precursor to plastic deformation and, in extreme cases, pressure-induced structural disorder or loss of long-range crystalline order. While the present analysis is limited to the harmonic elastic regime, elastic anisotropy is frequently associated with the early stages of mechanical instability preceding phase transitions or amorphization.

Importantly, the calculated bulk modulus shows excellent agreement with the independently calculated zero-pressure bulk modulus derived from the equation of state fit to the pressure–volume relationship, which yields a value of approximately 86 GPa. This close correspondence confirms the internal consistency of the elastic constant calculations and validates the reliability of both computational approaches.

Finally, the ratio $B/G$ = 2.45 provides insight into the mechanical behavior of the material. According to the empirical Pugh criterion,[42] materials with $B/G > 1.75$ are generally considered ductile, whereas lower values indicate brittleness. Therefore, the obtained value suggests that the material exhibits ductile behavior, implying a higher tendency to undergo plastic deformation before fracture. This ductility further supports the possibility that, under sufficient external pressure or stress, deformation may proceed through defect-mediated mechanisms rather than brittle failure, potentially contributing to progressive structural disorder.

## IV. Conclusions

In this work, the high-pressure structural behavior of β'-$Mn_3(PO_4)_2$ was investigated through a combination of synchrotron X-ray diffraction experiments and density-functional theory calculations. The results show that the monoclinic β' phase remains structurally stable up to approximately 14.1 GPa, where pronounced peak broadening and weakening in the XRD patterns reveal the onset of pressure-induced structural disorder. The persistence of these changes after decompression demonstrates that the transformation is irreversible and leads to a metastable disordered state rather than a crystalline high-pressure polymorph.

Compression of β'-$Mn_3(PO_4)_2$ is strongly anisotropic and is well described by a third-order Birch–Murnaghan equation of state. The experimentally determined bulk modulus, $K_0 = 81(2)$ GPa, indicates a compressibility comparable to other phosphate compounds and lower than that of related vanadates. Analysis of the compressibility tensor further reveals pronounced directional dependence, originating from the different roles played by rigid $PO_4$ tetrahedra and more deformable Mn–O polyhedra.

DFT calculations reproduce the compressional behavior determined from experiments and provide insight into local structural modifications under pressure. Compression is primarily accommodated through distortions of the Mn-centered polyhedra, while the $PO_4$ tetrahedra behave as nearly rigid units. Several initially penta-coordinated Mn sites progressively evolve toward octahedral coordination under pressure, while anomalous distortions and softening of selected $MnO_6$ polyhedra suggest the development of local structural instabilities before the onset of disorder.

The calculated elastic properties indicate that $\beta'$-$Mn_3(PO_4)_2$ is mechanically stable at ambient conditions but approaches elastic instability near 14 GPa, where one eigenvalue of the elastic tensor becomes negative, and the generalized Born stability criteria are no longer satisfied. The calculated elastic instability occurs near the pressure at which the diffraction pattern deteriorates. This correlation suggests that mechanical instability may contribute to the observed structural degradation, although the calculations do not uniquely identify the microscopic transformation pathway.

Overall, the combined experimental and computational results suggest that the high-pressure response of $\beta'$-$Mn_3(PO_4)_2$ is influenced by the interplay between framework topology, anisotropic polyhedral compressibility, and elastic stability. The comparison with previously studied $A_3(PO_4)_2$ phosphates indicates that structurally complex frameworks containing multiple crystallographically distinct cation sites and highly distorted coordination polyhedra may respond to compression differently from structurally simpler compounds, which can accommodate pressure through crystalline-to-crystalline phase transitions. Although additional studies on related phosphates are required to establish the generality of this behavior, the present work provides new insight into the compression mechanisms of structurally complex transition-metal phosphate frameworks and offers a useful basis for understanding the diverse high-pressure responses observed within this family of materials.

**Author contributions**

A. M. P. Brito, N. Bura, P. Botella, R. Oliva, A. K. Lopes da Costa, F. Vilella da Motta, M. R. D. Bomio, A. Muñoz, G. Garbarino, D. Errandonea: investigation (equal), methodology (equal), writing – review & editing (equal).

**Conflicts of interest**

The authors have no conflicts to disclose.

**Data availability**

The data that support the findings of this article are available from the authors upon reasonable request.

**Acknowledgements**

D.E. and A. M. thanks funding from Spanish Ministerio de Ciencia, Innovación y Universidades, through AEI (MCIN/AEI/10.13039/501100011033), and EU under grants PID2022-138076NB-C41/C44. P.B. and D.E. thank the financial support of Generalitat Valenciana through grant CIAPOS/2023/406. N.B. and D.E. thank the Generalitat Valenciana for the Postdoctoral Fellowship No. CIAPOS/2024/235. A.M.P.B. and M.B. are grateful for the financial support from the National Council for Scientific and Technological Development (CNPq)-Finance Code 403335/2023-0, and 304647/2024-1, respectively; and Coordination for the Improvement of Higher Education Personnel (CAPES)–Brazil with Finance Code 001. The authors thank the ESRF for providing beamtime for the experiments, proposal HC-6397.

**Table of contents entry**

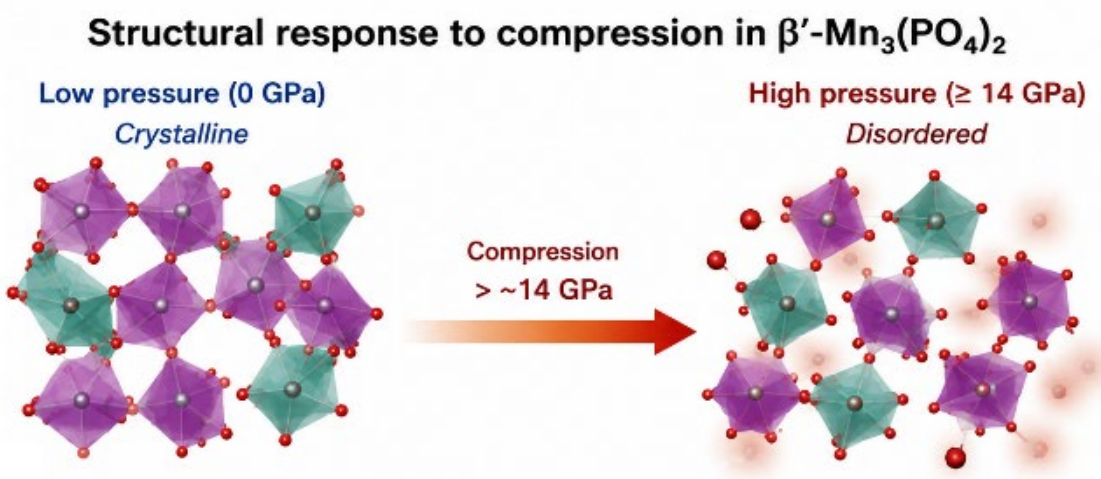

Framework complexity and elastic instability govern the high-pressure response of β'-$Mn_3(PO_4)_2$, leading to irreversible degradation of crystallinity near 14 GPa.